\documentstyle[twoside,fleqn,espcrc2,epsfig]{article}

\newcommand{\AmS}{{\protect\the\textfont2
  A\kern-.1667em\lower.5ex\hbox{M}\kern-.125emS}}
\newcommand{\ud}{\mathrm{d}}
\centerline {\bf ~~~~~~~~~~Study of Photonuclear Interaction of Muons in rock}
\centerline {\bf ~~~~~~~~~~with the MACRO experiment.}
\centerline {\bf ~~~~~~~~~~~~~~~~~E. Scapparone for the MACRO Collaboration}
\centerline{~~~~~~~~~~ INFN - Laboratori Nazionali del Gran Sasso}
\centerline{~~~~~~~~~~ S.S.17 km 18+910, 61070, Assergi (AQ), Italy}      
\centerline{ \bf ~~~~~~~~~~Talk at Xth International Symposium on Very High}
\centerline{ \bf ~~~~~~~~~~Energy Cosmic Ray Interactions}
\centerline{\bf ~~~~~~~~~~July 12 - 17, 1998}
\centerline{\bf ~~~~~~~~~~~~~To be published in Nucl. Phys. B, Proc. Suppl.~~~~~~~~~~}

\title{Study of Photonuclear Interaction of Muons in Rock \\
with the MACRO Experiment}

\author{E. Scapparone\address{INFN, LNGS
        S.S.17 km 18+910, 61070, Assergi (AQ), Italy}
       for the MACRO Collaboration\thanks{For the complete list of
the Collaboration see the paper "Relevance of the hadronic interaction
model in the interpretation of multiple muon data as detected with MACRO
experiment" by O. Palamara at these proceedings}
       }
\begin{document}

\begin{abstract}
We present first results about the measurement of the
charged hadrons production by atmospheric muons in
the rock above MACRO. 
A comparison between the measured rate with the Monte Carlo expectation
is presented.
\end{abstract}\vspace{0.2cm}

\maketitle
\section{Introduction}
The inelastic muon-nucleus interaction was 
studied at accelerators, 
in the range of large $Q^{2}$
transfer ($Q^{2}\ge 1 GeV^{2}$) 
mainly to measure
the nucleon structure functions (deep inelastic scattering experiments).
However, the bulk of interactions are characterized 
by low $Q^{2}$ ($Q^{2}\le 0.1 GeV^{2}$) and can be described with
the exchange of a quasi-real photon between the muon and the nucleon and they 
are often referred to as photo-nuclear interaction of muons.
Recently, it has been stressed that nuclear interactions of muons are an
important source of background for many underground experiments.
%
A comparison between experimental measurements about the hadrons production 
by muons and Monte Carlo (MC) simulations is mandatory to test the 
reliability of theoretical models.
In fact, uncertainties do exist on the cross section calculation for
the process, mainly due to the extrapolation of photo-nuclear 
cross section at high energies and in the 
simulation 
of the hadronic final state, because the smallness of the $Q^{2}$ does not 
allow the use of perturbative QCD.
The aim of the present work is to compare real data collected by the large area
underground experiment MACRO \cite{macro},
taking data at Gran Sasso laboratory, with GEANT 3.21 and
with FLUKA97, the latter based on the cross section calculated by Bezrukov and 
Bugaev\cite{bezrukov} and on DPM model\cite{dpm}
for the sampling of the final state. 
\section{The Detector}
A description of MACRO  
is given elsewhere\cite{macro}.  We remind here that MACRO operates at an average depth
of 3800 hg/cm$^2$, where the average residual energy of muons is
about 300 GeV. The
horizontal area of MACRO is about 1000 m$^2$ and charged tracks
can be reconstructed by means of a streamer tube system, 
in two projective views, with 
a space point accuracy of $\sim$1 cm.
The upper part of MACRO 
is in practice a thin detector, imposing a low threshold on secondary particles.
The lower part, thanks to the rock absorbers, allows to stop particles and 
to measure their
range up to few hundreds of MeV.
We consider events in which the 
muon interacts in the rock above the apparatus and both the muon and at least
one charged hadron are observed in the horizontal planes of
the tracking detector. An example is shown in Fig.1,
in which a muon enters the apparatus from above,
along with at least one charged hadron.
\begin{figure}[t]
\begin{center}
\begin{tabular}{c}
\mbox{\epsfig{file=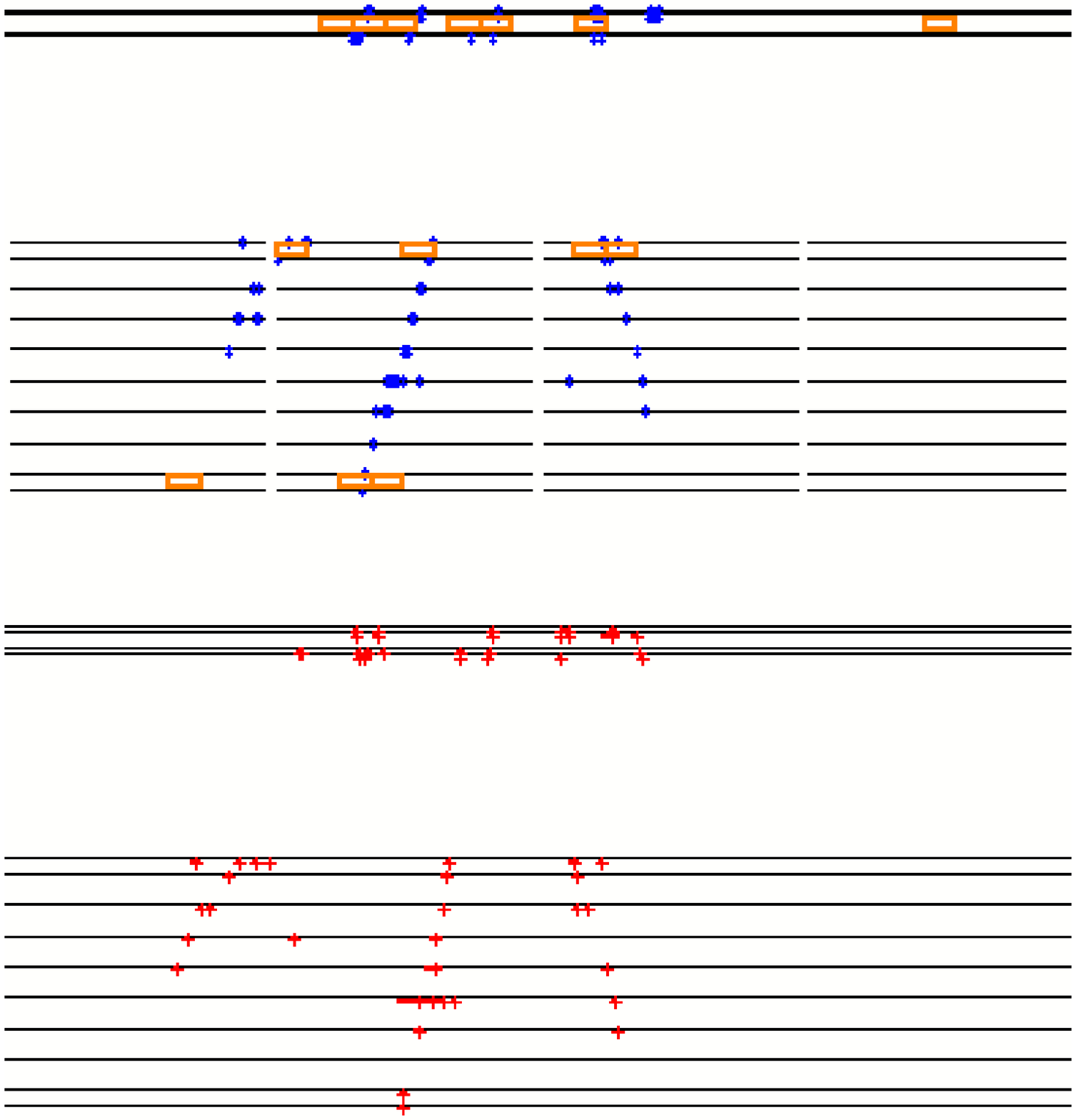,height=55mm,width=55mm}}\\
\mbox{\epsfig{file=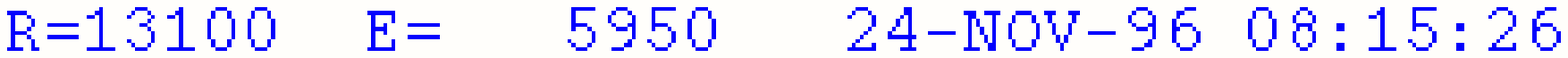,width=55mm}}\\
\end{tabular}
\vskip -0.5cm
\caption{\it A candidate event detected in MACRO.}
\label{bellevento}
\end{center}
\vskip -1cm
\end{figure}
Muon tagging is performed using the standard MACRO muon tracking while
the additional tracks have been recognized with a specialized
algorithm, looking for shorter tracks pointing towards the main track
around a common vertex region contained in the rock above the detector.
In practice we select charged hadrons with a minimum kinetic energy around 
150 MeV.
\section {The physical process and its simulation}
We have chosen FLUKA as a main reference to generate the underground muon events, 
taking into account the following steps.
The direction of incident muons is sampled according to the local angular 
distribution measured by MACRO and properly unfolded.
As far as muon residual energy at a depth $h$ (in km~w.e.) is 
concerned, we chose 
to sample it according to\cite{gaisser} 
\begin{equation}
\label{energy}
\frac{\ud N(E_\mu, h)}{\ud (E_\mu)} = K\, e^{-b\,h (\gamma - 1)}
\,(E_\mu + \epsilon \,(1-e^{-b\,h}))^{-\gamma}
\end{equation}
where $\gamma$=3.5,b=0.4(km\, w.e.$)^{-1}$ and
$\epsilon$ $\approx$540 GeV.

\begin{figure}[htb]
\vskip -1.2cm
\begin{center}
\vskip -0.1cm
\mbox{\epsfig{file=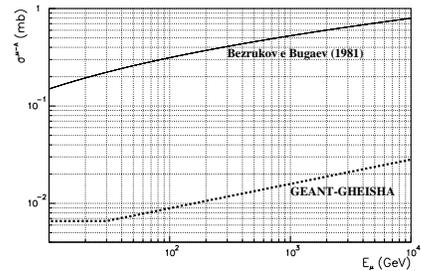,width=55mm}}
\vskip -0.5cm
\caption{\it The total muon photonuclear cross section in the considered models.}
\label{cross}
\end{center}
\vskip -1cm
\end{figure}

Given the direction, the slant depth of rock $h$ crossed by the muon can 
be obtained from the map of the mountain overburden.
In the simulation, muons are allowed to interact in a $13$~m thick layer of
rock positioned all around the experimental hall. This thickness corresponds to about 
\mbox{$35$ $\lambda_{int}$} for hadrons, and this is enough to fully
contain hadronic showers.
%
If a photo-nuclear interaction occurs in the region of rock described above,
the muon and secondary particles are transported through the rock, along with
e.m. and hadronic showers possibly produced. 
If both the muon and at least one additional particle reach the tunnel, the 
event is stored and a 
full simulation of the apparatus response is 
performed using the GEANT 3.21 package\cite{geant}.
The simulated data are treated with the
the same analysis instruments of real data.
In order to make a comparison with a different model, we have repeated
the event generation in the rock with GEANT.
One of the most important differences is in the total cross
section, as shown in Fig.~\ref{cross}.
For a more detailed discussion of the differences between FLUKA and GEANT-GHEISHA
for this process, see \cite{battist}.
\section { Data Analysis}
\begin{figure}[thb]
\begin{center}
\mbox{\epsfig{file=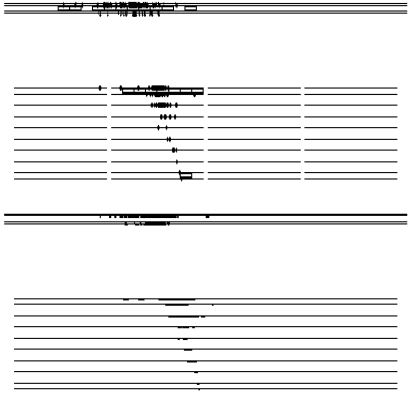,height=55mm,width=55mm}}
\vskip -0.5cm
\caption{\it Example of background event: e.m. interaction of muon
in the rock. \label{fi:back1}}
\end{center}
\vskip -1cm
\end{figure}

The main difficulty in our analysis is to achieve
the necessary rejection factor against the physical background.
Such a background is largely dominated by two processes: 
1) the e.m. interactions
of muons in the rock (bremsstrahlung and pair production, which
have a much larger cross section than the photonuclear reaction);
2) multiple muon events in which one of the muons stops inside
the detector.
An example of e.m. background event is shown in Fig.~\ref{fi:back1}.
\begin{figure}[thb]
\begin{center}
\vskip 0.5cm
\mbox{\epsfig{file=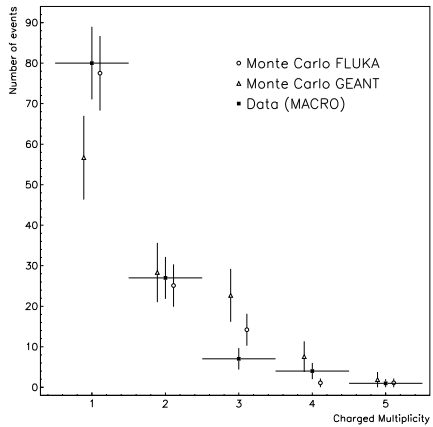,height=55mm,width=55mm}}
\vskip -2.cm
\caption{\it Charged hadrons multiplicity distribution}
\end{center}
\vskip -1cm
\end{figure}
This background has been extensively studied with our simulation
tools. In the case of the e.m. interactions, 
the mean angular separation between the muon and the additional tracks 
is less than the corresponding separation observed in photonuclear
interactions. Besides that, e.m. events often show very large clusters 
of fired tubes near the muon track.
These features are used to achieve a rejection factor against e.m. events
at the level of 4.5$\cdot$ 10$^{-6}$, 
maintaining the recognition efficiency for
hadronic events at the level of 55\%.
In order to reject the muon bundle backgrounds, additional cuts
based on the parallelism of tracks have to be considered.
We have studied the events generated with the HEMAS code\cite{hemas} 
using for the primary cosmic ray spectrum and mass composition
the results from the best fit reproducing the MACRO data 
themselves\cite{macro97b}
The rejection factor achieved
for muon bundle background at the level of 1.5$\cdot$10$^{-4}$,
at the price of a slight reduction of the selection efficiency,
which is now at 47\%.
\section { Results}
We have analysed a data sample corresponding
to about 11000 hours of full running of the detector.
With the above selection criteria, we have found
1938 candidate events over a total sample of
9544318 muon events.
From our knowledge of the background, we expect a
contamination by 11 events from the
e.m. interactions in the rock, and by 107 events muon bundles
surviving the cuts.
We can express the results in terms of the ratio 
$R_{\mu+h}$ of the selected $\mu$+hadrons events 
to the number of muon events in the same live time.
We then compare the experimental results to the MC prediction having
used the same selection criteria.
After the subtraction of the background, we find for $R_{\mu+h}$ 
in real data and in the MC simulations the following results:\\
\noindent- $R_{\mu+h}^{DATA}$~~~$=(1.91\pm0.05_{sta}\pm0.03_{sys})\cdot10^{-4}$,\\
- $R_{\mu+h}^{FLUKA}=(1.89\pm0.16_{sta}\pm0.02_{sys})\cdot10^{-4}$,\\ 
- $R_{\mu+h}^{GEANT}=(1.31\pm0.14_{sta}\pm0.02_{sys})\cdot10^{-5}$. \\
The systematic error on the experimental data is due to the uncertainties on
background subtraction, while in the simulation is
dominated by the uncertainties on the muon energy spectrum.
Fig. 4 shows the charged hadron multiplicty distribution. A good agreement
between the real data and FLUKA is obtained, while the GEANT distribution
is flatter, reflecting the $Q^{2}$ spectrum harder than FLUKA\cite{battist}.
From this preliminary measurement, we can conclude that the MACRO experiment 
can measure of the charged hadron (E$_{kin}$$>$150 MeV)
production in the rock at the desired level of accuracy.
The FLUKA predictions, based on the Bezrukov and Bugaev model,
are in very good agreement with data, while the GEANT-GHEISHA
model gives absolute predictions lower by an order of magnitude.
{ }
\end{document}